\documentclass[twocolumn,showpacs,preprintnumbers,amsmath,amssymb,showkeys]{revtex4}

\usepackage{epsfig}
\usepackage{amssymb}

\newcommand\eqn[1]{(\ref{#1})}      
\newcommand\Eqn[1]{Eq.~(\ref{#1})}  
\newcommand\Fig[1]{Fig.~\ref{#1}}  
\newcommand\Ref[1]{Ref.~\cite{#1}}  

\newcommand{\beq}{\begin{equation}}
\newcommand{\eeq}{\end{equation}}
\newcommand{\ba}{\begin{array}}
\newcommand{\bea}{\begin{eqnarray}}
\newcommand{\ea}{\end{array}}
\newcommand{\eea}{\end{eqnarray}}

\newcommand{\nn}{\nonumber\\}

\newcommand{\bx}{{\bf x}}
\newcommand{\by}{{\bf y}}
\newcommand{\br}{{\bf r}}
\newcommand{\bp}{{\bf p}}

\newcommand{\C}{\mathcal{C}}

\newcommand{\tr}{{\rm tr}}

\begin{document}


\title{Decoherence and thermalization of a pure quantum state \\in quantum field theory}

\author{Alexandre Giraud}
\author{Julien Serreau}
\affiliation{Astro-Particule et Cosmologie, Universit\'e Paris 7 -- Denis Diderot,\\ 10 rue A.~Domont et L.~Duquet, 75205 Paris cedex 13, France\footnote{APC is unit\'e mixte de recherche UMR7164 (CNRS, Universit\'e Paris 7, CEA, Observatoire de Paris)}}

\date{\today}

\begin{abstract}
We study the real-time evolution of a self-interacting $O(N)$ scalar field initially prepared in a pure, coherent quantum state. We present a complete solution of the nonequilibrium quantum dynamics from a $1/N$-expansion of the two-particle-irreducible effective action at next-to-leading order, which includes scattering and memory effects. We demonstrate that, restricting one's attention (or ability to measure) to a subset of the infinite hierarchy of correlation functions, one observes an effective loss of putity/coherence and, on longer time scales, thermalization. We point out that the physics of decoherence is well described by classical statistical field theory.
\end{abstract}

\maketitle

Quantum decoherence is a fundamental process whose understanding is a central issue in many areas of physics. Topical examples include measurement theory in quantum mechanics \cite{Zurek:1981xq}, the physics of Bose-Einstein condensates \cite{Graham}, or of quantum computers \cite{GPS}, neutrino physics \cite{Giunti:2002xg}, high-energy nuclear collisions \cite{Muller:2005yu}, black hole physics \cite{Hsu:2009ve}, or the description of primordial fluctuations in inflationary cosmology \cite{Polarski:1995jg, Campo:2008ju}. 
It is a genuine nonequilibrium process, which requires the real-time description of quantum dynamics. Analytic descriptions can be obtained for exactly solvable models and/or simple enough approximations, e.g. assuming a linear coupling between the system and its environment, neglecting back-reaction, etc. A complete microscopic description in realistic quantum field theories (QFT) is a notoriously difficult task \cite{Calzetta:1995ea}, which requires first principle calculations of the nonequilibrium quantum dynamics.

Recent years have witnessed substantial progress concerning the description of quantum fields out of equilibrium \cite{Berges:2004vw}. Two-particle-irreducible (2PI) functional techniques provide a powerful tool to devise systematic, practicable approximation schemes, valid for arbitrarily far from equilibrium situations \cite{Calzetta:1995ea,Berges:2004vw}. It has been demonstrated that a coupling or $1/N$-expansion of the 2PI effective action at lowest nontrivial order can describe far-from-equilibrium dynamics and subsequent (quantum) thermalization without further assumption \cite{Berges:2001fi, Aarts:2001yn,Berges:2000ur}. In this letter, we use 2PI techniques to compute the dynamics of decoherence from first principles in QFT \footnote{For recent works in a similar context, see \cite{Herranen:2008di, Koksma:2009wa}.}.

There are various uses of the concept of (de)coherence \cite{Anastopoulos:2000hg}, the most widely discussed being the so-called environment-induced decoherence, which results from the interaction of the system under study with a (thermal) bath of unobserved degrees of freedom (d.o.f.) \cite{Zurek:1981xq}. Even in the absence of an environment, decoherence of a subset of d.o.f. may result from some kind of coarse-graining \cite{Lombardo:1995fg, Habib:1995ee}. Here, we adopt a different point of view, also advocated e.g. in \cite{Calzetta:1995ea,Balian,Campo:2008ju, Koksma:2009wa}. Even if one keeps all the dynamical d.o.f., reconstructing the actual state of a given (closed) system requires a precise knowledge of its independent correlation functions. In practice, however, such information is often not experimentally accessible and one has to infer the state of the system from the subset of measured correlation functions. This ``incomplete description'' picture actually underlies the very concept of thermalization \cite{Balian,Calzetta:1995ea}. Similarly, a system prepared in a pure quantum state may appear as a statistical mixture to the observer who has only partial information. 

In this letter, we show that, starting from a pure, coherent quantum (Gaussian) state, an observer who only measures the subset of equal-time two-point functions, observes an effective loss of purity/coherence and, eventually, (apparent) thermalization. We study a relativistic self-interacting $O(N)$ scalar field and present a complete numerical solution of the nonequilibrium dynamics from a $1/N$-expansion of the 2PI effective action at next-to-leading order (NLO) \cite{Berges:2001fi}. The approach allows us to study the strong coupling regime and is a valid description for states with a high degree of quantum coherence which, as explained below,  are characterized by strong -- possibly non-perturbative -- field fluctuations. 

A pure quantum state remains such under a unitary evolution. Here, although the complete information about the initial Gaussian state is contained in the set of two-point functions, nontrivial higher correlators develop in time due to the non-Gaussian dynamics: The information about the initial state spreads in the space of correlation functions and gets lost to our observer, resulting in the effective loss of purity/coherence and, at late times, the thermalization of the effective density matrix. We stress that no environment of incoherent d.o.f. is needed. We illustrate this point by preparing a completely pure initial state. Finally, we show that the regime of decoherence, characterized by strong field fluctuations, is well-described by classical (statistical) field theory.

\begin{figure}[t!]
 \centerline{\epsfxsize=8.cm\epsffile{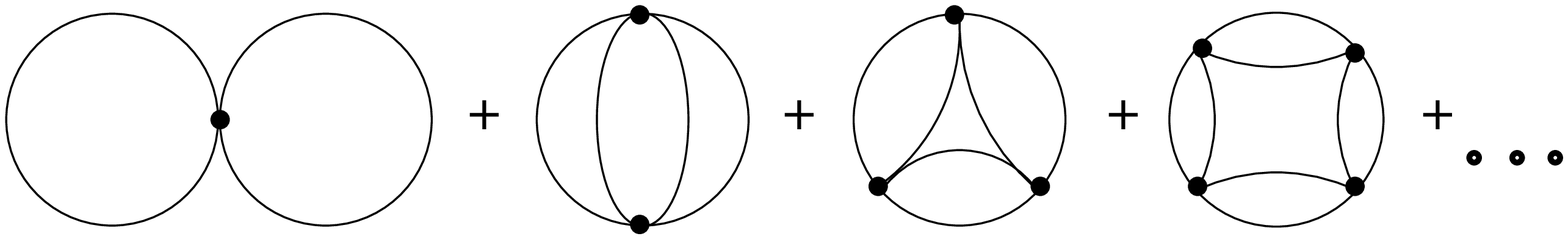}}
 \caption{\label{fig:SSBNLO} 
 The dots indicate that each diagram of the series is obtained from the previous 
 one by adding another bubble.\vspace*{-0.4cm}}
\end{figure}

We consider a relativistic real scalar field 
$\varphi_a$ ($a\!=\!1,\ldots, N$)
with classical action 
\beq
\label{eq:classical}
 S[\varphi]=-\!\! \int \!{\rm d}^4x \left\{ \frac{1}{2}\varphi_a\big(\square \!+\! m^2\big)\varphi_a\!+\!
\frac{\lambda}{4! N}\, \big(\varphi_a\varphi_a\big)^2 \right\}\,
\eeq 
where summation over repeated indices is implied.
For vanishing field expectation value, $\langle\varphi_a(x)\rangle=0$, the correlation functions of the quantum theory can be obtained from the 2PI effective action $\Gamma[G]$, parametrized by the time-ordered connected propagator 
$\langle {\rm T} \varphi_a(x)\varphi_b(y)\rangle =\delta_{ab}G(x,y)$ \cite{Cornwall:1974vz}. The $1/N$-expansion of $\Gamma[G]$ at NLO includes the infinite series of diagrams shown in \Fig{fig:SSBNLO}, where the lines represent the propagator $G$ \cite{Berges:2001fi}. 

To set up the initial-value problem, it is useful to decompose the two-point function $G$ into a statistical ($F$) and a spectral ($\rho$) component, both real functions:
\beq
 G(x,y)=F(x,y)-\frac{i}{2}{\rm sign}_\C(x^0-y^0)\rho(x,y)\,.
\eeq
Notice that $F(x,y)=F(y,x)$ and $\rho(x,y)=-\rho(y,x)$.
For Gaussian initial conditions, the equations of motion, obtained as $\delta \Gamma[G]/\delta G(x,y) = 0$, read
\bea
\left[\square_x +M^2(x)\right]F(x,y)\! &=&\!
- \int_{0}^{x^0}\!\!\!{\rm d}^4z\,\Sigma_{\rho}(x,z)F(z,y) \nonumber \\
&&\!+ \int_{0}^{y^0}\!\!\!{\rm d}^4z\, \Sigma_{F}(x,z)
\rho(z,y),
\label{eq:F1}
\\ 
\left[\square_x +M^2(x)\right]\rho(x,y)\! &=&\!
-\int_{y^0}^{x^0}\!\!\! {\rm d}^4z\,
\Sigma_{\rho}(x,z)\rho(z,y).
\label{eq:rho1}
\eea
where $\int_{0}^{x^0}\!\!\!{\rm d}^4z\equiv\int_{0}^{x^0}\!\!\!{\rm d}z^0\int \!{\rm d} {\bf z}$.
The effective mass term is given by 
\beq
M^2(x) = m^2 +\lambda\frac{N+2}{6N}F(x,x)
\eeq
and the self-energies by \cite{Berges:2001fi}
\bea
\hspace{-0.6cm}\Sigma_{F}(x,y) \!&=&\! \frac{\lambda}{3N} \Big[ F(x,y)I_{F}(x,y)
- \frac{1}{4} \rho(x,y) I_\rho(x,y) \Big], \!\!\!
\label{eq:sigmaF}\\
\hspace{-0.6cm}\Sigma_{\rho}(x,y) \!&=&\! \frac{\lambda}{3N}
\Big[\rho(x,y)I_{F}(x,y)+F(x,y)I_{\rho}(x,y)\Big]. \!\!\!
\label{eq:sigmarho}
\eea
The functions $I_{F}$ and $I_{\rho}$ resum the infinite series of bubble diagrams in \Fig{fig:SSBNLO}: 
\bea
I_{F}(x,y) = 
\Pi_{F}(x,y)
-\int_{0}^{x^0}\!\!\! d^4z\, I_{\rho}(x,z)\Pi_{F}(z,y)  
&&\nonumber \\
+\int_{0}^{y^0}\!\!\! d^4z\, I_{F}(x,z)\Pi_{\rho}(z,y)
&&\!\!\!\!\!\!\!, \label{eqIF1}\\
I_{\rho}(x,y) = \Pi_{\rho}(x,y)
-\int_{y^0}^{x^0}\!\!\! d^4z\,
I_{\rho}(x,z)\Pi_{\rho}(z,y),
&& \label{eq:Irho1}
\eea
with the elementary bubble
\bea
\Pi_{F}(x,y) &=& -\frac{\lambda}{6}\Big[F^2(x,y)-\frac{1}{4}\rho^2(x,y)\Big],
\label{eq:PIF}\\
\Pi_{\rho}(x,y) &=& -\frac{\lambda}{3}F(x,y)\rho(x,y).
\label{eq:PIR}
\eea

The self-energy kernels $\Sigma_F$ and $\Sigma_\rho$ describe scattering and memory effects and are responsible for non-Gaussian correlations to develop in time, a key ingredient in the present discussion. Gaussian -- collisionless -- evolution equations, such as leading-order Large-$N$, or Hartree approximations, where the RHSs of Eqs.~\eqn{eq:F1}-\eqn{eq:rho1} vanish identically, are known to fail to describe phenomena such as damping of unequal-time correlators or thermalization \cite{Berges:2004vw}. We show below that they also fail to capture the physics of decoherence in the present approach.

We consider spatially homogeneous and isotropic states, for which $F(x,y)\equiv F(x^0,y^0,|\bx-\by|)$. Accordingly, we introduce the Fourier decomposition ($p\equiv|\bp|$)
\beq
 F(t,t',|\br|)=\int \frac{{\rm d}{\bf p}}{(2\pi)^3}\,{\rm e}^{i\bp\cdot\br}\,F_p(t,t')\,,
\eeq
and similarly for $\rho(x,y)$. 
We consider an observer who only has access to the subset of equal-time two-point functions: $F_p(t)=\langle \varphi_\bp^\dagger(t)\varphi_\bp(t)\rangle$, $R_p(t)=\frac{1}{2}\langle \varphi_\bp^\dagger(t)\pi_\bp(t)+\pi_\bp^\dagger(t)\varphi_\bp(t)\rangle$ and $K_p(t)=\langle \pi_\bp^\dagger(t)\pi_\bp(t)\rangle$. The mimimum-bias state compatible with these measured observables is described by an effective Gaussian density matrix \cite{Balian} $D_{\rm eff}(t)=\prod_{\bp } D_{\bp }(t)$, where the product runs over independent d.o.f. in momentum space with, up to a normalization,
\bea
 &&\hspace{-.7cm}D_{\bp }(t)\propto\exp\left\{\kappa_p(t)\left[F_p(t)\pi_\bp^\dagger(t)\pi_\bp(t)+K_p(t)\varphi_\bp^\dagger(t)\varphi_\bp(t)\right.\right.\nn
 \label{eq:dm} &&\hspace{1.7cm}\left.\left.-R_p(t)(\pi_\bp^\dagger(t)\varphi_\bp(t)+\varphi_\bp^\dagger(t)\pi_\bp(t))\right]\right\},
\eea
where $\kappa_p(t)=-\ln[1+1/n_p(t)]/(2n_p(t)+1)$, with 
\beq
\label{eq:barn}
 n_p(t) +\frac{1}{2}=\sqrt{F_p(t)K_p(t)-R^2_p(t)\,}\equiv a_p(t)\,.
\eeq

Contrarily to the correlators $F_p(t)$, $R_p(t)$ and $K_p(t)$, which can be modified by a canonical redefinition of the field variables $(\varphi,\pi)$, $n_p(t)$ is a canonical invariant and, actually, the only truly intrinsic property of the density matrix \eqn{eq:dm} \cite{Campo:2008ju}. It provides an absolute measure of the quantum purity of the system's state through \footnote{The Schwartz inequality
$\langle A^\dagger A\rangle\langle B^\dagger B\rangle\ge|\langle A^\dagger B\rangle|^2$ with $A=\varphi_\bp(t)$ and $B=\pi_\bp(t)$, implies that $n_p(t)\ge0 \,\,\,\forall\,\, p,t$.}
\beq
\label{eq:trace}
 \tr\! \left[D_{\bp}^2(t)\right]=(2n_p(t)+1)^{-1}\le1\,,
\eeq
which equals $1$ for a pure state. Note also that, whenever the system is well-described by a thermal ensemble of weakly interacting quasi-particle (QP), $n_p(t)$ defines a QP occupation number and follows a Bose-Einstein distribution as a function of the QP energy \cite{Berges:2001fi}. We shall make use of this property to characterize the degree of thermalization of the system, although we stress that none of our results rely on such a QP interpretation. 

\begin{figure}[t!]
 \centerline{\epsfxsize=7.cm\epsffile{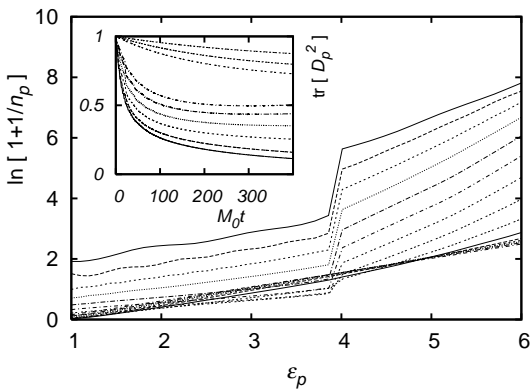}\vspace*{-0.2cm}}
 \caption{\label{fig:bose} 
 The function $\ln[1+1/n_p(t)]$ as a function of $\epsilon_p(t)$ for times $M_0t=2.5\times 2^n$ for $n=0,\dots,14$. A straight line at late times corresponds to a Bose-Einstein distribution. The inset shows the rapid decrease of $\tr[ D_p^2(t)]$ as a function of time for modes (from bottom to top) $p/a=1,\,5,\,10,\,15,\,20,\,24$ where $a/M_0=0.15$. Higher momentum modes, $p/a=25,\,27,\,30$, are also shown for comparison.\vspace*{-0.4cm}}
\end{figure}

Note finally that $n_p(t)$ is time-independent for Gaussian approximations \footnote{Gaussian dynamics is equivalent to a time-dependent canonical transformation; $n_p(t)$ is a canonical invariant.}. Therefore, no loss of quantum purity, in the sense discussed here, occurs for such approximations. We stress that this is not in conflict with existing studies of decoherence in free field \cite{Polarski:1995jg} or mean-field approximations \cite{Habib:1995ee}, which typically consider an effective coarse-grained Gaussian density matrix where one averages out some rapidly varying d.o.f. This results in an effective loss of information and hence of quantum purity/coherence, due e.g. to dephasing \cite{Habib:1995ee}, already at the Gaussian level. In the present incomplete description approach,  decoherence is due to the non-Gaussian dynamics, i.e. direct scattering and memory effects \footnote{The loss of information from the subset of Gaussian correlators is measured by the so-called correlation entropy \cite{Calzetta:1995ea,Balian,Campo:2008ju, Koksma:2009wa}: $S(t)=-\tr[D_{\rm eff}(t)\ln D_{\rm eff}(t)]=\sum_\bp[(n_p(t)+1)\ln(n_p(t)+1)-n_p(t)\ln n_p(t)]$.}.

We now come to the discussion of quantum coherence which, unlike purity, is a basis-dependent notion.
The equal-time two-point correlators can be parametrized as
\bea
\label{eq:init1} \epsilon_p(t)F_p(t)&=&\bar a_p(t)[1-\gamma_p(t)\cos\phi_p(t)]\,,\\
\label{eq:init2} R_p(t)&=&-\bar a_p(t)\gamma_p(t)\sin\phi_p(t)\,,\\
\label{eq:init3} K_p(t)/\epsilon_p(t)&=&\bar a_p(t)[1+\gamma_p(t)\cos\phi_p(t)]\,,,
\eea
where
\beq
\label{eq:n}
 \bar a_p(t)=\frac{K_p(t)+\epsilon^2_p(t)F_p(t)}{2\epsilon_p(t)}\equiv \bar n_p(t) +\frac{1}{2}
\eeq
provides an alternative definition of a QP occupation number $\bar n_p(t)$. Note that $0\le n_p(t)\le\bar n_p(t)$. Here, the energy scale $\epsilon_p(t)$ defines the QP basis in which to discuss decoherence. We use $\epsilon_p(t)=\sqrt{p^2+M^2(t)}$, which we found gives a good description of the oscillation frequency of two-point correlators. The basis-dependent occupation number $\bar n_p(t)$ is related to the canonical invariant $n_p(t)$ through the coherence parameter \footnote{Note that $\bar n_p(t)$ and $\gamma_p(t)$, not being canonical invariants, are in general not conserved by Gaussian dynamics.}
\beq
\label{eq:gamma}
\gamma_p(t)=\sqrt{1-a^2_p(t)/\bar a^2_p(t)} \,.
\eeq
The case $\gamma_p(t)=0$ or, equivalently, $n_p(t)= \bar n_p(t)$, corresponds to the thermal-like density matrix $D_\bp(t)\propto\exp\{\kappa_p(t) F_p(t)\left[\pi_\bp^\dagger(t)\pi_\bp(t)+\epsilon_p^2(t)\varphi_\bp^\dagger(t)\varphi_\bp(t)\right]\}$. This includes the vacuum-like state: $n_p(t)=\gamma_p(t)=0$. Moreover, $\gamma_p(t)$ controls the size of off-diagonal matrix elements of the density matrix in the so-called two-mode coherent state basis \footnote{It can be directly related to the parameter $\delta$ of \Ref{Campo:2008ju}.}: The latter exhibits non-trivial correlations (i.e. quantum coherence) between macroscopically distant semi-classical states for $\gamma_p(t)\to 1$. Note that this limit is also characterized by strong field fluctuations since the correlators $F_p(t)$, $R_p(t)$ and $K_p(t)$ are $\propto1/\sqrt{1-{\gamma_p}^{\!\!2}(t)}\gg1$. 

\begin{figure}[t!]
 \centerline{\epsfxsize=7.5cm\epsffile{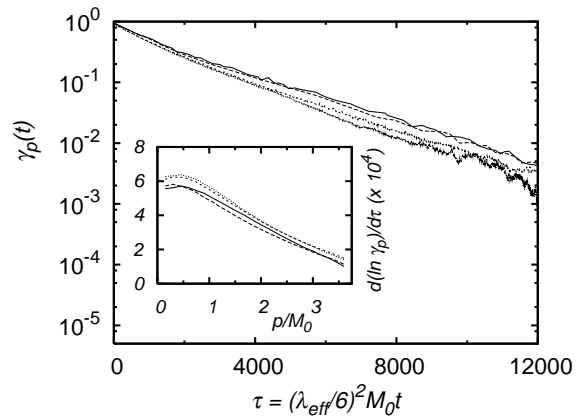}\vspace*{-.2cm}}
 \caption{\label{fig:gamma} 
 Exponential decay of the coherence parameter $\gamma_p(t)$ as a function of rescaled time $\tau=(\lambda_{eff}/6)^2(M_0t)$ (see text), for the mode $p/M_0=0.15$, for four couples of parameters ($\gamma_0,\lambda$): ($0.99,3.05$) [solid] and ($0.95,6.62$) [long-dashed], for which $\lambda_{eff}=21.7$, ($0.95,2.2$) [dashed], for which $\lambda_{eff}=7.215$ and ($0.95,0.915$) [dotted], for which $\lambda_{eff}=3.0$. The inset shows the corresponding decay rates, obtained from exponential fits, for modes $p\le p_c$. The agreement of the first two runs illustrates the  $\lambda_{eff}$-scaling of early-time decoherence. The overall agreement shows the approximate $\lambda_{eff}^2$-dependence of the decoherence rate.\vspace*{-0.4cm} }
\end{figure}

At time $t=0$, we prepare modes with $p\le p_c$ in a pure, highly coherent quantum state characterized by $n_p(0)\ll1$, $\gamma_p(0)=\gamma_0\sim 1$ and $\phi_p(0)=\pi/2$, while modes with $p_c<p\le \Lambda$, with $\Lambda$ the ultra-violet cutoff, are prepared in a vacuum-like state: $n_p(0)\ll1$ and $\gamma_p(0)=0$ \footnote{Initial conditions for the spectral correlator $\rho$ are fixed by the equal-time commutation relations of field operators.}. We emphasize that there are no incoherent d.o.f.: The initial state is a completely pure quantum state. Eqs.~\eqn{eq:F1}-\eqn{eq:PIR} are solved numerically without further approximation (except for discretization of time/momentum-integrals and truncation of memory-in\-tegrals). We show results for $p_c/M_0=3.6$ and $\Lambda/M_0=10$, where $M_0=M(0)$ is the initial effective mass.

\Fig{fig:bose} presents an overview of the time-evolution. It shows snapshots of the function $\ln[1+1/n_p(t)]$ as a function of $\epsilon_p(t)$ at various times. One observes a substantial growth of $n_p(t)$ for all modes, signaling the effective loss of quantum purity at early times, followed by a slow approach to an effective quantum thermal equilibrium, characterized by a Bose-Einstein distribution. The loss of quantum purity is further illustrated in the inset, which shows the rapid decay of ${\rm tr}[D_p^2(t)]$ for modes $p\le p_c$.

After the loss of quantum purity, we find that $\bar n_p(t)\approx n_p(t)$, signaling a corresponding loss of quantum coherence according to \Eqn{eq:gamma}. \Fig{fig:gamma} shows the time evolution of the coherence parameter $\gamma_p(t)$ for the mode $p/M_0=0.15$, for various sets of parameters. In all cases we find that decoherence is well-described by an exponential law. The inset shows the corresponding decoherence rates for the modes $p\le p_c$. 

{\it Classical scaling:} In the highly coherent limit,  $\gamma_0\to1$, one has, roughly speaking $F\sim1/\sqrt{1-\gamma_0^2}\gg\rho\sim1$ which signals the enhancement of classical {\it vs.} quantum fluctuations \cite{Aarts:2001yn}. In this regime one can neglect the second term in brackets on the LHSs of Eqs.~\eqn{eq:PIF} and \eqn{eq:sigmaF}. It is then easy to check that under the simultaneous rescaling of the correlators and the coupling constant: $F\to\eta F$, $\rho\to\rho$ and $\lambda\to\lambda/\eta$, with $\eta$ an arbitrary constant, the $F$ and $\rho$-components of $\Pi$, $I$ and $\Sigma$ scale just as $F$ and $\rho$ respectively and that Eqs.~\eqn{eq:F1}-\eqn{eq:rho1} are left invariant. This is characteristic of the regime of strong field fluctuations, where a classical (statistical) field theory description is appropriate. Indeed, in classical field theory, the action being defined up to a multiplicative constant, a rescaling of the field (i.e. of the initial conditions) can be entirely absorbed in a change of coupling. Therefore, as long as the system is highly coherent ($\gamma_p(t)\sim1$), we expect the dynamics not to depend separately on $\gamma_0$ and $\lambda$, but instead on the combination $\lambda_{eff}=\lambda/\eta$ where $\eta=\sqrt{1-\gamma_0^2}$. We checked, from our numerical simulation of the full quantum dynamics, that this is indeed the case during the early-time decoherence regime, as illustrated in \Fig{fig:gamma}: Runs with different values of $\gamma_0$ and $\lambda$ but the same $\lambda_{eff}$ are essentially indistinguishable. Furthermore, we observe that, despite the rather strong (effective) couplings employed in some simulations, the decoherence rates approximately follow a perturbative-like $\lambda_{eff}^2$-scaling, as also illustrated on \Fig{fig:gamma}.

As a final remark, we mention that fixing the value of the initial mass $M_0$ absorbs a large $\Lambda^2$-dependence. This simple, though approximate, renormalization turns out to be numerically sufficient for the results presented here. We find that, if the late time thermalization is cutoff-dependent, a known artifact of Gaussian initial conditions \cite{Borsanyi:2008ar}, the regime of effective loss of quantum purity/coherence is largely cut-off insensitive.

To the best of our knowledge, this work provides the first complete microscopic description of the process of decoherence in a realistic QFT. It is an exciting observation that the relevant dynamics is well-described by classical statistical field theory, which can be solved exactly by means of standard Monte-Carlo Techniques. A particularly interesting question is to investigate how the present results are modified as one includes the knowledge of higher order equal-time correlation functions in the set of measured observables. 

{\it Acknowledgements:} We thank R. Balian for interesting discussions and for pointing out \Ref{Balian}  as well as D.~Campo and R.~Parentani for interesting discussions and useful suggestions concerning the manuscript.

\end{document}